\documentclass[8pt]{article}
\usepackage[utf8]{inputenc}
\usepackage{amsmath}
\RequirePackage{fix-cm} % arbitrary font scaling
\DeclareMathSizes {8} {8} {7} {4}
\usepackage{multicol}
\usepackage[tiny]{titlesec}
\usepackage{graphicx}
\graphicspath{ {./images/} }
\begin{document}
\begin{center}
\large{Optimization of Modified Quantum Discord in Projector Space}
\end{center}
\begin{center}
\small{Problem 1 : A Size Invariant NP-Complete Optimization, '16 - '17}
\end{center}
\begin{center}
    \small{Chitradeep Gupta, IIT Delhi}
\end{center}
\begin{abstract}
In a pair of correlated quantum systems a measurement in one corresponds to a change in the state of the other. In the process, information about the original state of the system is lost. Measurement along which set of projectors would accompany minimum loss in information content is the optimization problem of quantum discord and is an important aspect of a classical to quantum transition because it asks us to look for the most classical states. This optimization problem is known to be NP-complete and is important because discord is defined through it making it a major obstacle on every computation. The standard zero discord condition helps us move to a stronger measure that addresses the correlated observables instead, in such a context we show that minimum discord occurs at the diagonal basis of the reduced density matrices and present an analytical expression of the measure. The work employs manipulations of information inequalities that leads to an exact optimization.

\end{abstract}

\section{Introduction}
Quantum Discord is the primary indicator of quantum correlations[1] in information theory and has been thoroughly studied over the past two decades. The work is about finding an explicit connection between discord and commutation, how information loss through discord becomes breakdown of commutation in terms of the geometry in Hilbert space. We study the formalism in search for an explicit connection between the two and reach the standard zero discord condition to find quantum superpositions that still linger on. We show that removing these correlations at zero discord takes us to a stronger measure in terms of correlated observables. This helps us go around the NP-complete optimization[2] that discord is defined by. Thus this is also an addition to P vs NP literature. The optimization problem itself is meaningful in how it asks us to look for the projectors of the most classical states that would give rise to minimum loss in information content as we make a measurement. Understanding discord is therefore critical to the classical to quantum transition and should deepen our reach further in quantum foundations. The formal development of quantum discord opens with the insight that mutual information $I(S:A)$ between two systems S and A that is symmetric in a classical setting fails to be in a quantum context with ${J(S:A)}_{\{{\Pi}^{A}_{j}\}}$ where ${\{{\Pi}^{A}_{j}\}}$ is the set of projectors in A along which the first measurement is made as we study how S behaves in return.
\[ I(S:A) = H(A) + H(S) - H(S,A)\]
\[{J(S:A)}_{\{{\Pi}^{A}_{j}\}} = H(S)-H(S|A)_{\{{\Pi}^{A}_{j}\}}\]
The difference between the two measures gives us discord $\delta(S:A)_ {\{{\Pi}^{A}_{j}\}}$ for a specific set of projectors that in turn speaks of the information loss as we move from the pre- to the post-measurement states. Discord between S and A is then defined through an optimization over the entire projector space of A and this is known to be intractable[2].
\begin{align*}
\delta(S:A)_ {\{{\Pi}^{A}_{j}\}} 
&= I(S:A)-{J(S:A)}_{\{{\Pi}^{A}_{j}\}}\\
&= H(A) + H(S|A)_{\{{\Pi}^{A}_{j}\}} - H(S,A)\\
\delta(S:A) &= min(\delta(S:A)_ {\{{\Pi}^{A}_{j}\}}) 
\end{align*}
Analytical computations[3-6] for this optimization problem is done but a concrete theorem on this problem is not known. Along which set of projectors we should make the measurement for minimum information loss is crucial for any quantum measurement and therefore, we approach this problem from information inequalities and show that this optimization with an extra set of projectors always occurs at the diagonal basis of the reduced density matrices for a compound density matrix that is separable. We had to design a strictly larger variable based on contradictions from principles so that the minimization could happen to both the variables together and we could go around Huang's theorem[2] on the problem being NP-complete to reach an analytical expression of a measure of quantum discord. In the expression being symmetric in S and A at the minima even when for any arbitrary set of projectors it is not is where we get to see how commutativity can arise at the most classical projection. We also show how taking any projection in Hilbert space is accompanied by information loss which minimizes at the diagonal basis.\\
\\
\textbf{A Language of Correlated Observables :} We would like to begin with the zero discord condition[1], $\rho_{\scriptscriptstyle{S,A}} = \rho_{\scriptscriptstyle{S,A}}^{\scriptscriptstyle{D}} = \sum_{j}\Pi^{\scriptscriptstyle{A}}_{j}\rho_{\scriptscriptstyle{S,A}}\Pi^{\scriptscriptstyle{A}}_{j}$  , it is the equivalence of the pre-measurement and post-measurement states. A $\rho_{\scriptscriptstyle{S,A}}$ of the form $\sum_{j}P_{j}\rho^{j}_{\scriptscriptstyle{S}}\times \Pi^{\scriptscriptstyle{A}}_{j}$ would satisfy such a condition. With a spectral decomposition we would arrive at the relation,
\[ \rho_{\scriptscriptstyle{S,A}} = \sum_{i,j}\Pi^{\scriptscriptstyle{S}}_{j,i}\Pi^{\scriptscriptstyle{A}}_{j} \rho_{\scriptscriptstyle{S,A}} \Pi^{\scriptscriptstyle{A}}_{j}\Pi^{\scriptscriptstyle{S}}_{j,i}\]
Zero discord is equivalent to finding the set of measurements $\{\Pi^{\scriptscriptstyle{A}}_{j}, \Pi^{\scriptscriptstyle{S}}_{j,i}\}$ for which the density matrix is not disturbed and no information is lost for any other observer unaware of the outcomes. However, the set of measurements $\{\Pi^{\scriptscriptstyle{S}}_{j,i}\}$ in S for a given $j$ is dependent on the outcome of the measurement in A. Here we would like to make an observation,\\
\\
\textbf{Observation 1} In the zero discord condition for $\delta(S:A)$ what outcome one gets for A determines which observable one needs to measure in S.\\
\\
From the perspective of dynamics this can lead to very absurd grounds because the outcome of say spin up on the first electron means we will measure the momentum of the second electron and if it is spin down we will measure energy in a scenario where energy and momentum do not commute. This choice of basis is not there in classical theory and if we have to understand the classical to quantum transition from quantum discord we need an extra set of projectors so that these quantum correlations that are still left at zero discord is removed. A more classical measure for correlations therefore, should answer the question of how much the distribution of a certain particular observable in S changes based on the outcome in A which leads us to define the mutual information ${J(S:A)}_{\{{\Pi}^{A}_{j},{\Pi}^{S}_{i}\}}$,
\[{J(S:A)}_{\{{\Pi}^{A}_{j},{\Pi}^{S}_{i}\}} = H(S) - H(S|A)_{\{{\Pi}^{A}_{j},{\Pi}^{S}_{i}\}}\]
$H(S|A)_{\{{\Pi}^{A}_{j},{\Pi}^{S}_{i}\}}$ is defined as $H(\rho^{\scriptscriptstyle{D}}_{\scriptscriptstyle{{S|\{\Pi^A_{j}\}}}})_{\{\Pi^S_{i}\}}$. While there is gain in information about S upon conditioning with A by $H(\rho_{\scriptscriptstyle{S}}) - H(\rho_{\scriptscriptstyle{S}|\{\Pi^A_{j}\}})$ there is loss in information about S upon projection along $\{\Pi^S_{i}\}$ by $H(\rho_{\scriptscriptstyle{S}|\{\Pi^A_{j}\}}) - H(\rho^{\scriptscriptstyle{D}}_{\scriptscriptstyle{{S|\{\Pi^A_{j}\}}}})_{\{\Pi^S_{i}\}}$. ${J(S:A)}_{\{{\Pi}^{A}_{j},{\Pi}^{S}_{i}\}}$ is then the sum of both expressing the amount of information gained about a certain particular observable in S from a measurement of a certain other observable in A. The extra information that is lost upon projection is exclusively a quantum phenomena which would go to zero in the classical case. This establishes the otherwise classical equivalence of the two mutual informations ${J(S:A)}_{\{{\Pi}^{A}_{j},{\Pi}^{S}_{i}\}}$ and $I(S:A)$ which aligns with the original starting point of Ollivier and Zurek in 2001. In such a setting one could define a discord $\alpha(S:A)$ as follows,   
\begin{align*}
    \alpha(S:A)_{\{{\Pi}^{A}_{j},{\Pi}^{S}_{i}\}} 
    &= I(S:A)-{J(S:A)}_{\{{\Pi}^{A}_{j},{\Pi}^{S}_{i}\}}\\
    &= H(A)+H(S|A)_{\{{\Pi}^{A}_{j},{\Pi}^{S}_{i}\}}-H(S,A)\\
    \alpha(S:A) &= min(\alpha(S:A)_{\{{\Pi}^{A}_{j},{\Pi}^{S}_{i}\}})
\end{align*}
$\alpha(S:A)_{\{{\Pi}^{A}_{j},{\Pi}^{S}_{i}\}}$ is the information loss when a series of successive measurements is made of two particular observables in the respective Hilbert Spaces of A, and then in S. $\alpha(S:A)$ is defined by the optimization of that over the two projector spaces in A and then S. Notice that one of the intentions behind doing this is to move to a larger measure than the standard version so that the minimization stays relevant and we could conclude on a stronger zero discord.

\section{Optimization of Modified Quantum Discord in Projector Space} 
Minimization of $\alpha$ depends on the minimization of $H(S|A)_{\{{\Pi}^{A}_{j},{\Pi}^{S}_{i}\}}$. At this point we would like to introduce a lemma, trying to understand what this term means explicitly. The Hilbert spaces of S and A are spanned by the states $\{|n_{i}>\}$ and $\{| k_{j}>\}$. The notation $P(x,y)$ stands for the probability of some state $|x>$ to collapse at $|y>$.\\
\\
\textbf{Lemma 1} ${H(S|A)}_{\{{\Pi}^{A}_{j},{\Pi}^{S}_{i}\}} = H(\rho^{\scriptscriptstyle{D}}_{\scriptscriptstyle{{S,A}}})_{\{\Pi^S_{i}\times\Pi^A_{j}\}} - H(\rho^{\scriptscriptstyle{D}}_{\scriptscriptstyle{{A}}})_{\{\Pi^A_{j}\}} = D(P(SA,n_{i}k_{j})||P(A,k_{j}))$\\
\\
\textit{Proof:} This can be proved straightforwardly by chasing the summations. On the third line we prove the relation on relative entropy.
\begin{align*}
H(\rho^{\scriptscriptstyle{D}}_{\scriptscriptstyle{{S|\{\Pi^A_{j}\}}}})_{\{\Pi^S_{i}\}} &= \sum_{j}P(A,k_{j})H(\rho^{\scriptscriptstyle{D}}_{\scriptscriptstyle{{S|\Pi^A_{j}}}})_{\{\Pi^S_{i}\}}\\
&= -\sum_{j,i}P(A,k_{j})P(S,n_{i}|A,k_{j})log(P(S,n_{i}|A,k_{j}))\\
&= -\sum_{j,i}P(SA,n_{i}k_{j})log(P(SA,n_{i}k_{j})/P(A,k_{j}))\\
&= -\sum_{j,i}P(SA,n_{i}k_{j})log(P(SA,n_{i}k_{j})) + \sum_{j,i}P(SA,n_{i}k_{j})log(P(A,k_{j}))\\
&= -\sum_{j,i}P(SA,n_{i}k_{j})log(P(SA,n_{i}k_{j})) + \sum_{j}P(A,k_{j})log(P(A,k_{j}))\\
&=  H(\rho^{\scriptscriptstyle{D}}_{\scriptscriptstyle{{S,A}}})_{\{\Pi^S_{i}\times\Pi^A_{j}\}} - H(\rho^{\scriptscriptstyle{D}}_{\scriptscriptstyle{{A}}})_{\{\Pi^A_{j}\}}
\end{align*}
The next lemma helps us establish that any form of measurement or projection in the Hilbert space is always accompanied with a loss in information content. The loss is zero only when the projection is along the diagonal basis.\\
\\
\textbf{Lemma 2} $H(\rho^{\scriptscriptstyle{D}}_{\{{\Pi_{j}\}}}) \geq H(\rho)$\\
\\
\textit{Proof:} $\{\lambda_{i}\}$ are the eigenvalues of $\rho$ with eigenstates $\{\Bar{k}_{i}\}$ while projection is taken along $\{k_{j}\}$. We apply Jensen's inequality[7] to complete the proof.
\begin{align*}
    H(\rho^{\scriptscriptstyle{D}}_{\{{\Pi_{j}\}}}) &= H(\sum_{j,i}\Pi_{j}\lambda_{i}\Bar{\Pi}_{i}\Pi_{j})\\
    &= H(\sum_{j,i}P(k_{j},\Bar{k}_{i})\lambda_{i}\Pi_{j})\\
    &= -\sum_{j}(\sum_{i}P(k_{j},\Bar{k}_{i})\lambda_{i})log(\sum_{i}P(k_{j},\Bar{k}_{i})\lambda_{i})\\
    &\geq -\sum_{j,i}P(k_{j},\Bar{k}_{i})\lambda_{i}log\lambda_{i}\\
    &\geq -\sum_{i}\lambda_{i}log\lambda_{i}\\
    &\geq  H(\rho)
\end{align*}
\textbf{Theorem 1} $\alpha(S:A)_{\{{\Pi}^{A}_{j},{\Pi}^{S}_{i}\}}$ for a separable density matrix $\rho_{\scriptscriptstyle{S,A}}$ minimizes along the projectors of the diagonal basis of its reduced density matrices.\\
\\
\textit{Proof:} The density matrix we choose to work with is a mixture of otherwise classical density matrices of the form $\sum_{j}P_{j}\Pi^{\scriptscriptstyle{S}}_{j} \times \Pi^{\scriptscriptstyle{A}}_{j}$. The resultant matrix is of the form,
\[ \rho_{\scriptscriptstyle{S,A}} = \sum_{i,j}P_{i}P^{i}_{j}\Pi^{\scriptscriptstyle{S}}_{i,j} \times \Pi^{\scriptscriptstyle{A}}_{i,j}\] 
In this case, the minimization of $\alpha$ by lemma 1 is just that of $D(P(SA,n_{i'}k_{j'}||P(A,k_{j'}))$, as in $D(P_{i'j'}||P_{j'}))$. The reduced density matrices have eigenbasis $\{\Bar{{\Pi}}^{A}_{j''}\}$ and $\{\Bar{{\Pi}}^{S}_{i''}\}$. 
\begin{align*}
P_{i'j'} &= \sum_{i,j}P_{i}P^{i}_{j}P(n_{ij},n_{i'})P(k_{ij},k_{j'})  \tag{1}\\ 
     \rho_{\scriptscriptstyle{A}} &= \sum_{i,j}P_{i}P^{i}_{j}\Pi^{\scriptscriptstyle{A}}_{i,j} = \sum_{j''}\Bar{P}_{j''}\Bar{\Pi}^{\scriptscriptstyle{A}}_{j''} \\
 \Bar{P}_{j''} &= \sum_{i,j}P_{i}P^{i}_{j}P(k_{ij},\Bar{k}_{j''}) \tag{2}\\
    P_{j'} &= \sum_{i,j}P_{i}P^{i}_{j}P(k_{ij},k_{j'}) \tag{3}\\ 
    P_{j'} &= \sum_{j''}\Bar{P}_{j''}P(\Bar{k}_{j''},k_{j'}) \\
    &= \sum_{i,j,j''}P_{i}P^{i}_{j}P(k_{ij},\Bar{k}_{j''})P(\Bar{k}_{j''},k_{j'}) \tag{4}
\end{align*}
We arrived at (4) using (2). Comparing (4) and (3) gives us the following equations,
\begin{align*}
P(k_{ij},k_{j'}) &= \sum_{j''}P(k_{ij},\Bar{k}_{j''})P(\Bar{k}_{j''},k_{j'}) \tag{5}\\
P(n_{ij},n_{i'}) &= \sum_{i''}P(n_{ij},\Bar{n}_{i''})P(\Bar{n}_{i''},n_{i'}) \tag{6}
\end{align*}
It is important to notice here that the chain rule multiplication of probabilities in (5) and (6) followed by a summation over the exclusive alternatives wouldn't work for any other basis other than $\{|\Bar{n}_{i''}>\}$ or $\{|\Bar{k}_{j''}>\}$ because of the interference terms[8]. Equations (5) and (6) ask us to recognize that the states of S and A are essentially a classical mixture only in their diagonal basis and no other and it is this realization that helps us minimize discord.\\
\\
We use equations (1) and (3) to find ${H(S|A)}_{\{{\Pi}^{A}_{j'},{\Pi}^{S}_{i'}\}}$ from lemma 1, introduce (5) at the next step and employ log-sum inequality[9] with respect to the index $\{j''\}$. The $j'$-terms within the logarithm cancel out and a summation over $\{j'\}$ gives us the minimization at ${\{\Bar{{\Pi}}^{A}_{j''}}\}$.
\begin{align*}
{H(S|A)}_{\{{\Pi}^{A}_{j'},{\Pi}^{S}_{i'}\}} &= -\sum_{i',j'}P_{i'j'}log(\frac{P_{i'j'}}{P_{j'}})\\
&= -\sum_{i',j'}({ \sum_{i,j}P_{i}P^{i}_{j}P(n_{ij},n_{i'})P(k_{ij},k_{j'})})log(\frac{ \sum_{i,j}P_{i}P^{i}_{j}P(n_{ij},n_{i'})P(k_{ij},k_{j'})}{\sum_{i,j}P_{i}P^{i}_{j}P(k_{ij},{k}_{j'})})\\
&= -\sum_{i',j'}({ \sum_{i,j,j''}P_{i}P^{i}_{j}P(n_{ij},n_{i'})P(k_{ij},\Bar{k}_{j''})P(\Bar{k}_{j''},k_{j'})})log(\frac{\sum_{i,j,j''}P_{i}P^{i}_{j}P(n_{ij},n_{i'})P(k_{ij},\Bar{k}_{j''})P(\Bar{k}_{j''},k_{j'})}{\sum_{i,j,j''}P_{i}P^{i}_{j}P(k_{ij},\Bar{k}_{j''})P(\Bar{k}_{j''},k_{j'})})\\
&\geq -\sum_{i',j',j''}({ \sum_{i,j}P_{i}P^{i}_{j}P(n_{ij},n_{i'})P(k_{ij},\Bar{k}_{j''})P(\Bar{k}_{j''},k_{j'})})log(\frac{\sum_{i,j}P_{i}P^{i}_{j}P(n_{ij},n_{i'})P(k_{ij},\Bar{k}_{j''})P(\Bar{k}_{j''},k_{j'})}{\sum_{i,j}P_{i}P^{i}_{j}P(k_{ij},\Bar{k}_{j''})P(\Bar{k}_{j''},k_{j'})})\\
&\geq -\sum_{i',j''}({ \sum_{i,j}P_{i}P^{i}_{j}P(n_{ij},n_{i'})P(k_{ij},\Bar{k}_{j''})})log(\frac{\sum_{i,j}P_{i}P^{i}_{j}P(n_{ij},n_{i'})P(k_{ij},\Bar{k}_{j''})}{\sum_{i,j}P_{i}P^{i}_{j}P(k_{ij},\Bar{k}_{j''})})\\
&\geq {H(S|A)}_{\{\Bar{{\Pi}}^{A}_{j''},{\Pi}^{S}_{i'}\}}
\end{align*}
We use equations (1) and (3) to find ${H(S|A)}_{\{\Bar{{\Pi}}^{A}_{j''},{\Pi}^{S}_{i'}\}}$ from lemma 1, introduce (6) at the next step and employ Jensen's inequality for concave functions[7] with respect to the distribution $\{P(\Bar{n}_{i''},n_{i'})\}$ with the varying index as $\{{i''}\}$. A summation over $\{i'\}$ then gives us the minimization at ${\{\Bar{{\Pi}}^{S}_{i''}}\}$ and completes the proof.
\begin{align*}
    {H(S|A)}_{\{\Bar{{\Pi}}^{A}_{j''},{\Pi}^{S}_{i'}\}} &= -\sum_{i',j''}({ \sum_{i,j}P_{i}P^{i}_{j}P(n_{ij},n_{i'})P(k_{ij},\Bar{k}_{j''})})log(\frac{\sum_{i,j}P_{i}P^{i}_{j}P(n_{ij},n_{i'})P(k_{ij},\Bar{k}_{j''})}{\sum_{i,j}P_{i}P^{i}_{j}P(k_{ij},\Bar{k}_{j''})})\\
    &= -\sum_{i',j''}({ \sum_{i,j,i''}P_{i}P^{i}_{j}P(n_{ij},\Bar{n}_{i''})P(\Bar{n}_{i''},n_{i'})P(k_{ij},\Bar{k}_{j''})})log(\frac{\sum_{i,j,i''}P_{i}P^{i}_{j}P(n_{ij},\Bar{n}_{i''})P(\Bar{n}_{i''},n_{i'})P(k_{ij},\Bar{k}_{j''})}{\sum_{i,j}P_{i}P^{i}_{j}P(k_{ij},\Bar{k}_{j''})})\\
    &\geq -\sum_{i',j'',i''}P(\Bar{n}_{i''},n_{i'})({\sum_{i,j}P_{i}P^{i}_{j}P(n_{ij},\Bar{n}_{i''})P(k_{ij},\Bar{k}_{j''})})log(\frac{\sum_{i,j}P_{i}P^{i}_{j}P(n_{ij},\Bar{n}_{i''})P(k_{ij},\Bar{k}_{j''})}{\sum_{i,j}P_{i}P^{i}_{j}P(k_{ij},\Bar{k}_{j''})})\\
    &\geq -\sum_{i'',j''}({\sum_{i,j}P_{i}P^{i}_{j}P(n_{ij},\Bar{n}_{i''})P(k_{ij},\Bar{k}_{j''})})log(\frac{\sum_{i,j}P_{i}P^{i}_{j}P(n_{ij},\Bar{n}_{i''})P(k_{ij},\Bar{k}_{j''})}{\sum_{i,j}P_{i}P^{i}_{j}P(k_{ij},\Bar{k}_{j''})})\\
    &\geq {H(S|A)}_{\{\Bar{{\Pi}}^{A}_{j''},\Bar{{\Pi}}^{S}_{i''}\}}
\end{align*}
The most general $m \times n$ separable density matrix $\sum_{i}P_{i}\rho^{i}_{\scriptscriptstyle{S}} \times \rho^{i}_{\scriptscriptstyle{A}}$ upon decomposition would give $\sum_{i,j,l}P_{i}P^{i}_{j}P^{i}_{l}\Pi^{\scriptscriptstyle{S}}_{i,j} \times \Pi^{\scriptscriptstyle{A}}_{i,l}$. The index $l$ is being ignored for simplification. The reader is encouraged to check how the proof still gets through.\\
\\
\textbf{Theorem 2} $\alpha(S:A)$ for a separable density matrix $\rho_{\scriptscriptstyle{S,A}}$ is symmetric in S and A.\\
\\
\textit{Proof:} $\alpha(S:A) = \alpha(S:A)_{\{\Bar{{\Pi}}^{A}_{j},\Bar{{\Pi}}^{S}_{i}\}}\\
= H(A)+H(S|A)_{\{\Bar{\Pi}^{A}_{j},\Bar{\Pi}^{S}_{i}\}}-H(S,A)\\
= H(A) + H(\rho^{\scriptscriptstyle{D}}_{\scriptscriptstyle{{S,A}}})_{\{\Bar{\Pi}^S_{i} \times \Bar{\Pi}^A_{j}\}} - H(\rho^{\scriptscriptstyle{D}}_{\scriptscriptstyle{{A}}})_{\{\Bar{\Pi}^A_{j}\}} - H(S,A)\\
= H(\rho^{\scriptscriptstyle{D}}_{\scriptscriptstyle{{S,A}}})_{\{\Bar{\Pi}^S_{i}\times \Bar{\Pi}^A_{j}\}} - H(S,A)$\\
\\
The symmetry that broke down at $J(S:A)_{\{{\Pi}^{A}_{j},{\Pi}^{S}_{i}\}}$ is again restored in $\alpha(S:A)$ while giving us an analytical expression of discord in equation (7). 
\begin{align*}
\alpha(S:A) &= H(\rho^{\scriptscriptstyle{D}}_{\scriptscriptstyle{{S,A}}})_{\{\Bar{\Pi}^S_{i}\times \Bar{\Pi}^A_{j}\}} - H(S,A) \tag{7}
\end{align*}
This is the expression we had strived for, notice how the projectors in equation (7) is a function of the given density matrix. The symmetry that arises at the point of minima with the most classical states shows how the sequence of measurements in the classical realm is inconsequential and how commutivity arises at the most classical projection. Equation (7) and lemma 2 together helps us attain quite a few key insights that we would like to discuss at length.

\section{Implications}

(i) \textbf{The zero discord condition :} $\alpha(S:A)=0$ can only occur when the set of projectors ${\{\Bar{{\Pi}}^{S}_{i} \times \Bar{{\Pi}}^{A}_{j}\}}$ becomes the diagonal basis of the given density matrix, there by restricting it to the form $\sum_{i,j}P_{ij}{{\Pi}}^{S}_{i}\times\Pi^{A}_{j}$ with product eigenstates. In other words, unless $[\rho_{\scriptscriptstyle{{S,A}}},N \times K]=0$ for some N and K in the respective Hilbert spaces of S and A, information loss through discord is inevitable. This shows that non-commutivity is the underlying cause of nonzero discord and establishes the explicit connection we looked for. This gives it a firm standing on quantum foundations.\\
\\
(ii) \textbf{The strongness condition :} $\alpha(S:A)_{\{{{\Pi}}^{A}_{j},{{\Pi}}^{S}_{i}\}} \geq \delta(S:A)_{{\{{\Pi}}^{A}_{j}\}}$. This is a direct consequence of lemma 2 since $\alpha$ only differs from $\delta$ by an extra projection. The zero discord condition of $\alpha(S:A)$ in (i) assigns both the discords $\delta(S:A)$ and $\delta(A:S)$ to be zero together. This shows that the new measure we defined is a stronger measure of quantum discord. However, it is important to understand that the underlying intentions of both the formalisms are different. The seminal work of Ollivier and Zurek[1] was an attempt at understanding the measurement problem in QM in terms of quantum information. We try to see the connection in foundations and where is QI aligned with QM.\\
\\
(iii) \textbf{Identification of Pointer States :} The minimization of $\alpha(S:A)_{\{{\Pi}^{A}_{j},{\Pi}^{S}_{i}\}}$ shows that the diagonal basis of the reduced density matrices are a strong candidate for Pointer States. This is in alignment with lemma 2 where we saw that measurement along the diagonal basis doesn't lead to any information loss and hence, are most classical. Equation (7) in that sense, talks about the closest one could get to such a scenario in the product space of two correlated systems where the diagonal basis of the joint density matrix is essentially entangled for a nonzero discord. 

\section{Instances and Discussions} 
A comparative study of both the discords is being done as we see it play out for real systems.\\
\\
{\textit{\underline{$\alpha$ for the Werner states} \textbf{$\rho_{w}= \frac{1}{4}(I + x\Vec{\sigma_{1}}.\Vec{\sigma_{2}})$}}}

\begin{multicols}{2}
\includegraphics[width=7cm, height=5.5cm]{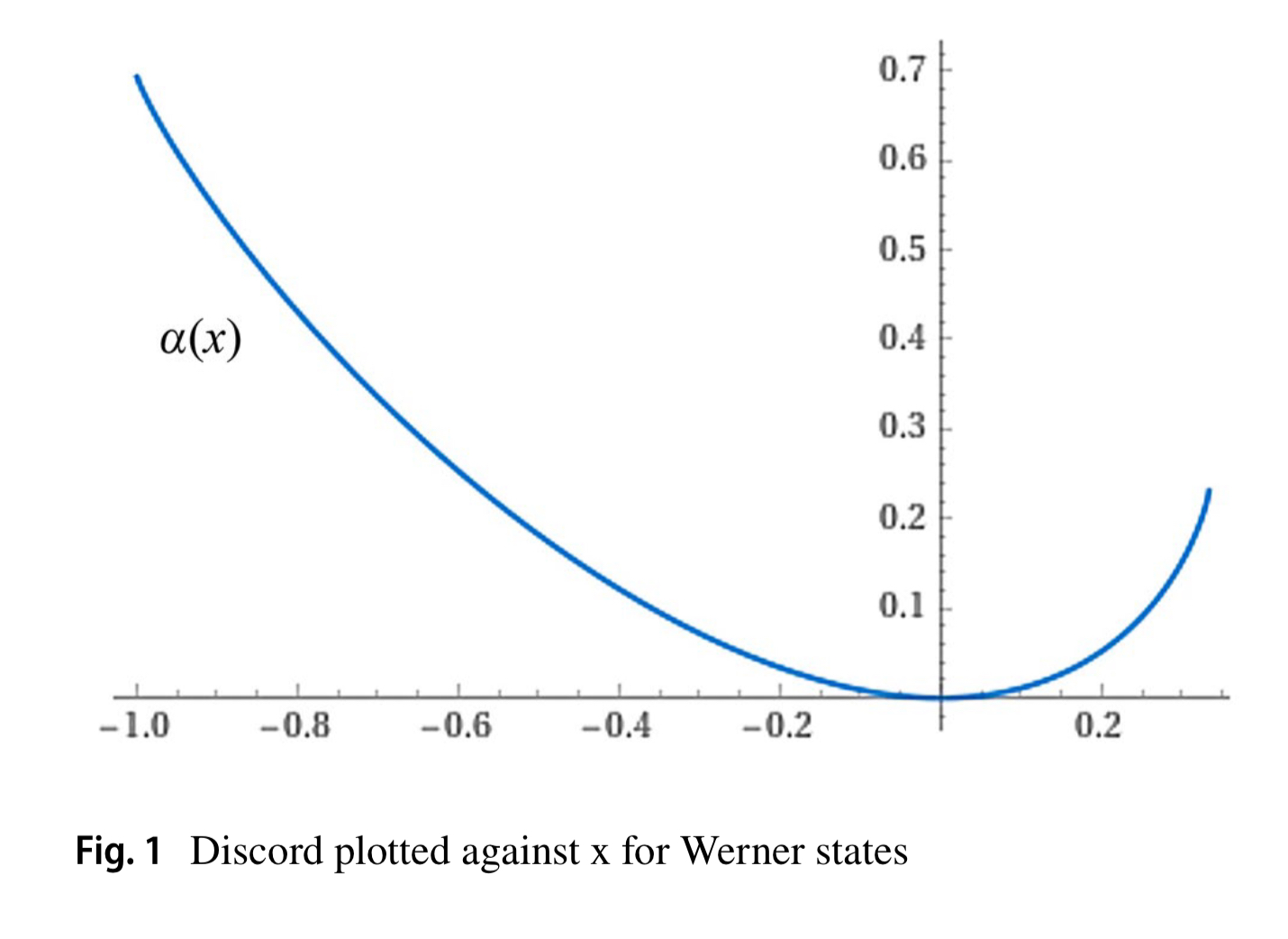}
\\
\[\alpha(x)= \frac{1+x}{4}log\frac{1+x}{4} + \frac{1-3x}{4}log\frac{1-3x}{4} - \frac{1-x}{2}log\frac{1-x}{4}\]
The curve for $\alpha(x)$ that we have obtained behaves exactly as $\delta(x)$ does in Werner states as demonstrated in [10].
\end{multicols}

\textit{\underline{$\alpha$ for the example in Zurek'01}} 
\textbf{$\rho_{z}= \frac{1}{2}(|00><00| + |11><11|) + \frac{z}{2}(|00><11| + |11><00|)$}\\
\includegraphics[width=15cm, height=6cm]{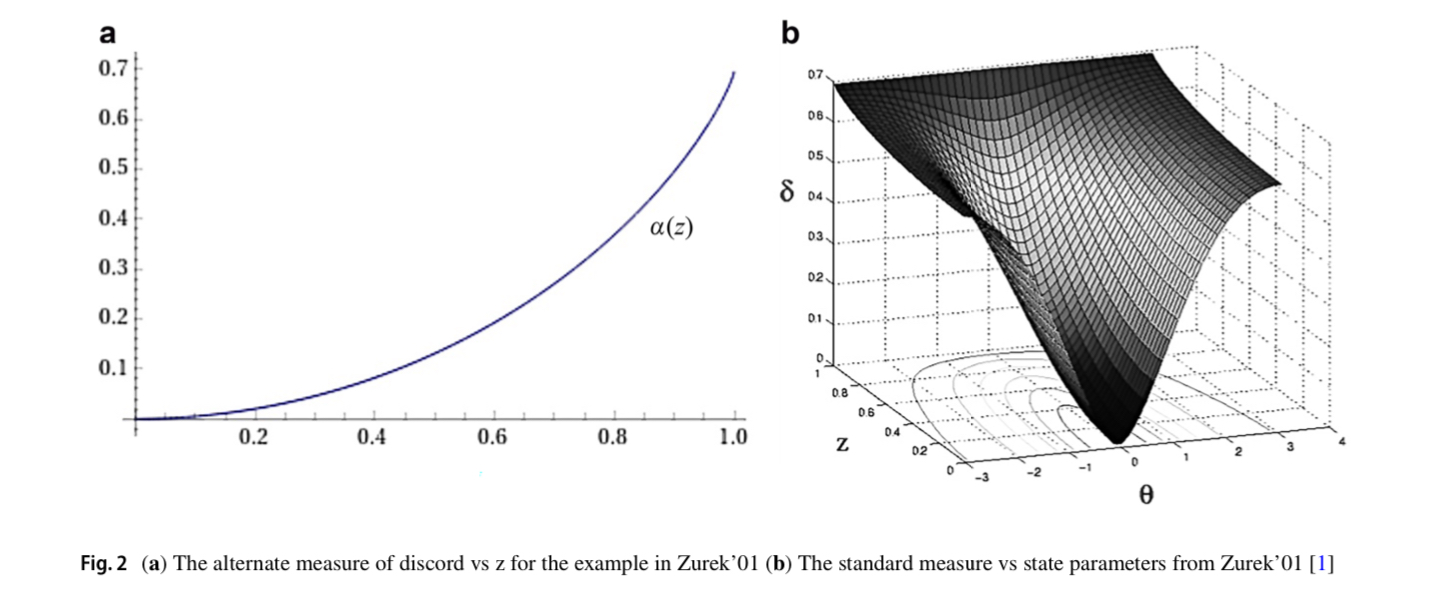}
\[\alpha(z)= \frac{1+z}{2}log\frac{1+z}{2} + \frac{1-z}{2}log\frac{1-z}{2} + log2\]

The curve for $\alpha$ that we have obtained for the original example in [1] corresponds to the spine of the plot given in Zurek'01[1]. The variation over the projector space by $\theta$ and how discord minimizes at the spine follows from Theorem 1. So it seems $\alpha$ and $\delta$ agree fairly well. For asymmetric states in S and A, they of course would not for contributions from both $\delta(S:A)$ and $\delta(A:S)$ play into $\alpha$ because of the symmetry at minima. While $\delta(S:A)$ was originally designed by Ollivier and Zurek[1] to address the measurement problem from the perspective of information loss in a state S when measured by an apparatus A, $\alpha(S:A)$ is designed to address two correlated states S and A.

\section{Conclusion} 
This work has been firmly grounded to understanding quantum discord in terms of quantum foundations with an attempt at finding a complete picture of the classical to quantum transition[11] in terms of information theory. The results on the NP-complete optimization in projector space happened with time. A stronger measure for discord got formalized to address the quantumness in correlations between two systems. The optimization problem is then solved in Theorem 1 for a separable $m \times n$ matrix to produce an analytical expression of a discord at Theorem 2 that turns out to be symmetric at minimum discord. This helps us to identify the most classical states (pointer states) as the diagonal basis of the reduced density matrices that when measured would disturb the systems the least with minimum loss in information content. That in terms of correlations, the diagonal basis of the two states are least correlated quantum mechanically. Notice that this proposition is in alignment with the symmetry that emerges upon optimization because the classical realm is commutative in terms of projection. In terms of P vs NP, we do not believe we have resolved the issue but we learned to deal with an important NP-complete problem and we believe this could be helpful. This being the synopsis of our main results, one does wonder about the exact scope of Theorem 1. We chose to work with a separable density matrix because it is known now that entanglement is not necessary for a nonzero discord[1]. Having said that the proof heavily relies on the separability of $\rho_{S,A}$. If such a theorem cannot be established for entangled states at all then one could devise a new separability criterion from it which we find highly unlikely. The confidence in Theorem 1 is reinforced via equation (7) which establishes a strong equivalence between information loss through discord and the breakdown of commutation in geometry in sect. 3-(i).
\section{ Acknowledgement} 
The author thanks Prof. V. Ravishankar for giving us his valuable time and helpful comments.
\section{References}
$[1]$ Ollivier, H. and Zurek, W. (2001) "Quantum Discord: A Measure of the Quantumness of Correlations" Phys. Rev. Lett. 88, 017901\\
\\
$[2]$ Huang, Y. (2014) "Computing quantum discord is NP-complete"  New Journal of Physics 16, 033027\\
\\
$[3]$ Ali, M. Rau, A. and Alber, G. (2010) "Quantum discord for two-qubit X states" PRA, 81(4):042105\\
\\
$[4]$ Girolami, D. and Adesso, G. "Quantum discord for general two-qubit states: Analytical progress" PRA-83, 052108\\
\\
$[5]$ Vinjanampathy, S. and Rau, A. (2012) "Quantum discord for qubit–qudit systems" J. of Phys. A: Mathematical and Theoretical\\
\\
$[6]$ Lu, X. Ma, J. Xi, Z. and Wang, X. (2011) "Optimal measurements to access classical correlations of two-qubit states" Phys. Rev. A, 83(1):012327\\
\\
$[7]$ Cover, T. and Thomas, J. (1991) "Elements of Information Theory" Chapter-16, Theorem-16.1.2\\
\\
$[8]$ Feynman, R. (1948) "Space-Time Approach to Non-Relativistic Quantum Mechanics" Rev.Mod.Phys. 20, 367\\
\\
$[9]$ Cover, T. and Thomas, J. (1991) "Elements of Information Theory" Chapter-16, Theorem-16.1.1\\
\\
$[10]$  Díaz-Solórzano, S. and Castro, E. (2018) "Exact analytical solution of Entanglement of Formation and Quantum Discord for Werner state and Generalized Werner-Like states" arXiv:1802.05877 [quant-ph], Figure-2\\
\\
$[11]$ Zurek, W. (2003). ”Decoherence, einselection and the quantum origins of the classical”. RMP-75 (3): 715–775

\end{document}